\documentclass[envcountsect]{llncs}

\usepackage{times}
\usepackage{amssymb}
\usepackage{amsmath}
\usepackage{latexsym}
\usepackage{graphics}
\usepackage{color}
\usepackage{url}

\spnewtheorem{fact}{Fact}[section]{\bfseries}{\itshape}
\spnewtheorem{observation}{Observation}[section]{\bfseries}{\itshape}

\newenvironment{myparagraph}
{\setlength{\parskip}{0in}

\begin{list}{$\bullet$}%
         {\setlength{\leftmargin}{40pt}
         \setlength{\topsep}{0pt}
         \setlength{\parskip}{0in}
         \setlength{\parsep}{0in}
         \setlength{\itemsep}{0in}}%

} {\end{list}}

\begin{document}
\title{ A certifying algorithm for 3-colorability of $P_5$-free
graphs}
\author{
Daniel Bruce \thanks{Computing and Information Science, University
of Guelph, Canada. \texttt{ email: dbruce01@uoguelph.ca}}  \ \ \
\ \ Ch\'{\i}nh T. Ho\`{a}ng\thanks{Physics and Computer Science,
Wilfred Laurier University, Canada.  Research supported by NSERC.
\texttt{ email: choang@wlu.ca}} \ \ \  \ \ Joe
Sawada\thanks{Computing and Information Science, University of
Guelph, Canada. Research supported by NSERC.  \texttt{ email:
jsawada@uoguelph.ca}} }

\institute{}

\date{\today}
\maketitle
\begin{abstract}
We provide a certifying algorithm for the problem of deciding
whether a $P_5$-free graph is 3-colorable by showing there are
exactly six finite graphs that are $P_5$-free and not 3-colorable
and minimal with respect to this property.
\end{abstract}

\section{Introduction}
An algorithm is {\it certifying} if it returns with each output a
simple and easily verifiable certificate that the particular
output is correct. For example, a certifying algorithm for the
bipartite graph recognition would return either a 2-coloring of
the input graph proving that it is bipartite, or an odd cycle
proving it is not bipartite. A certifying algorithm for planarity
would return a planar embedding or one of the two Kuratowski
subgraphs. The notion of certifying algorithm \cite{KraMcc} was
developed when researchers noticed that a well known planarity
testing program was incorrectly implemented. A certifying
algorithm is a desirable tool to guard against incorrect
implementation of a particular algorithm. In this paper, we give a
certifying algorithm for the problem of deciding whether a
$P_5$-free graph is 3-colorable. We will now discuss the
background of this problem.

A class {\cal C} of graphs is called {\it hereditary} if for each
graph $G$ in {\cal C}, all induced subgraphs of $G$ are also in
${\cal C}$. Every hereditary class of graphs can be described by
its \emph{forbidden induced subgraphs}, i.e. the unique set of
minimal graphs which do not belong to the class. A comprehensive
survey on coloring of graphs in hereditary classes can be found in
\cite{RS04}. An important line of research on colorability of
graphs in hereditary classes deals with $P_t$-free graphs. The
induced path on $t$ vertices is called $P_t$, and a graph is
called \emph{$P_t$-free} if it does not contain $P_t$ as an
induced subgraph.

It is known  \cite{woe}  that $5$\textsc{-colorability} is
NP-complete for $P_8$-free graphs and  \cite{LRS06}
$4$\textsc{-colorability} is NP-complete for $P_9$-free graphs. On
the other hand, the $k$-{\sc colorability} problem can be solved
in polynomial time for $P_4$-free graphs (since they are perfect).
In \cite{Hoa} and \cite{Hoa2}, it is  shown that
$k$\textsc{-colorability} can be solved for the class of
$P_5$-free graphs in polynomial time for every particular value of
$k$. For $t=6,7$, the complexity of the problem is generally
unknown, except for the case of $3$\textsc{-colorability} of
$P_6$-free graphs \cite{RS04A}. Known results on the
$k$\textsc{-colorability} problem in $P_t$-free graphs are
summarized in Table \ref{table} ($n$ is the number of vertices in
the input graph, $m$ the number of edges, and $\alpha$ is matrix
multiplication exponent known to satisfy $2 \leq \alpha < 2.376$
\cite{cop}).

\begin{table}\label{table}
\begin{center}
\begin{tabular}{|c|c|c|c|c|c|c|c|c|c|c|c|}
\hline
$k\backslash t$&3&4&5&6&7&8&9&10&11&12&\ldots\\
\hline
3 &$O(m)$ &$O(m)$ &$O(n^\alpha)$ &$O(mn^\alpha)$&?&?&?&?&?&?&\ldots\\
4 &$O(m)$ &$O(m)$ &P  &? &? &?      & $NP_c$ & $NP_c$& $NP_c$& $NP_c$ &\ldots\\
5 &$O(m)$ &$O(m)$ &P  &? &? &$NP_c$ & $NP_c$ & $NP_c$& $NP_c$& $NP_c$ &\ldots\\
6 &$O(m)$ &$O(m)$ &P  &? &? &$NP_c$ & $NP_c$ & $NP_c$& $NP_c$& $NP_c$ &\ldots\\
7 &$O(m)$ &$O(m)$ &P  &? &? &$NP_c$ & $NP_c$ & $NP_c$& $NP_c$& $NP_c$ &\ldots\\
\ldots&\ldots&\ldots&\bf{\ldots}&\ldots&\ldots&\ldots&\ldots&\ldots&\ldots&\ldots&\ldots\\
\hline
\end{tabular}

\smallskip

\caption{Known complexities for $k$-colorability of $P_t$-free
graphs}
\end{center}
\end{table}

In this paper, we study the coloring problem for the class of
$P_5$-free graphs. This class has proved resistant with respect to
other graph problems. For instance, $P_5$-free graphs is the
unique minimal class defined by a single forbidden induced
subgraph with unknown complexity of the {\sc maximum independent
set} and {\sc minimum independent dominating set} problems. Many
algorithmic problems are known to be NP-hard in the class of
$P_5$-free graphs, for example {\sc dominating set} \cite{Kor90}
and {\sc chromatic number} \cite{kral}. In contrast to the
NP-hardness of finding the chromatic number of a $P_5$-free graph,
it is known \cite{Hoa} that $k$\textsc{-colorability} can be
solved in this class in polynomial time for every particular value
of $k$. This algorithm produces a $k$-coloring if one exists, but
does not produce an easily verifiable certificate when such
coloring does not exist. We are interested in finding a
certificate for non-$k$-colorability of $P_5$-free graphs. For
this purpose, we start with $k=3$.

Besides \cite{Hoa}, there are several polynomial-time algorithms
for 3-coloring a $P_5$-free graph (\cite{Hoa2,mel,woe}) but none
of them is a certifying algorithm. In this paper, we obtain a
certifying algorithm for 3-coloring a $P_5$-free graphs by proving
there are a finite number of minimally non-3-colorable $P_5$-free
graphs and each of these graphs is finite.

\begin{figure}[ht]
\begin{minipage}[b]{0.5\linewidth}
\centering \resizebox{!}{2.0in}{\includegraphics{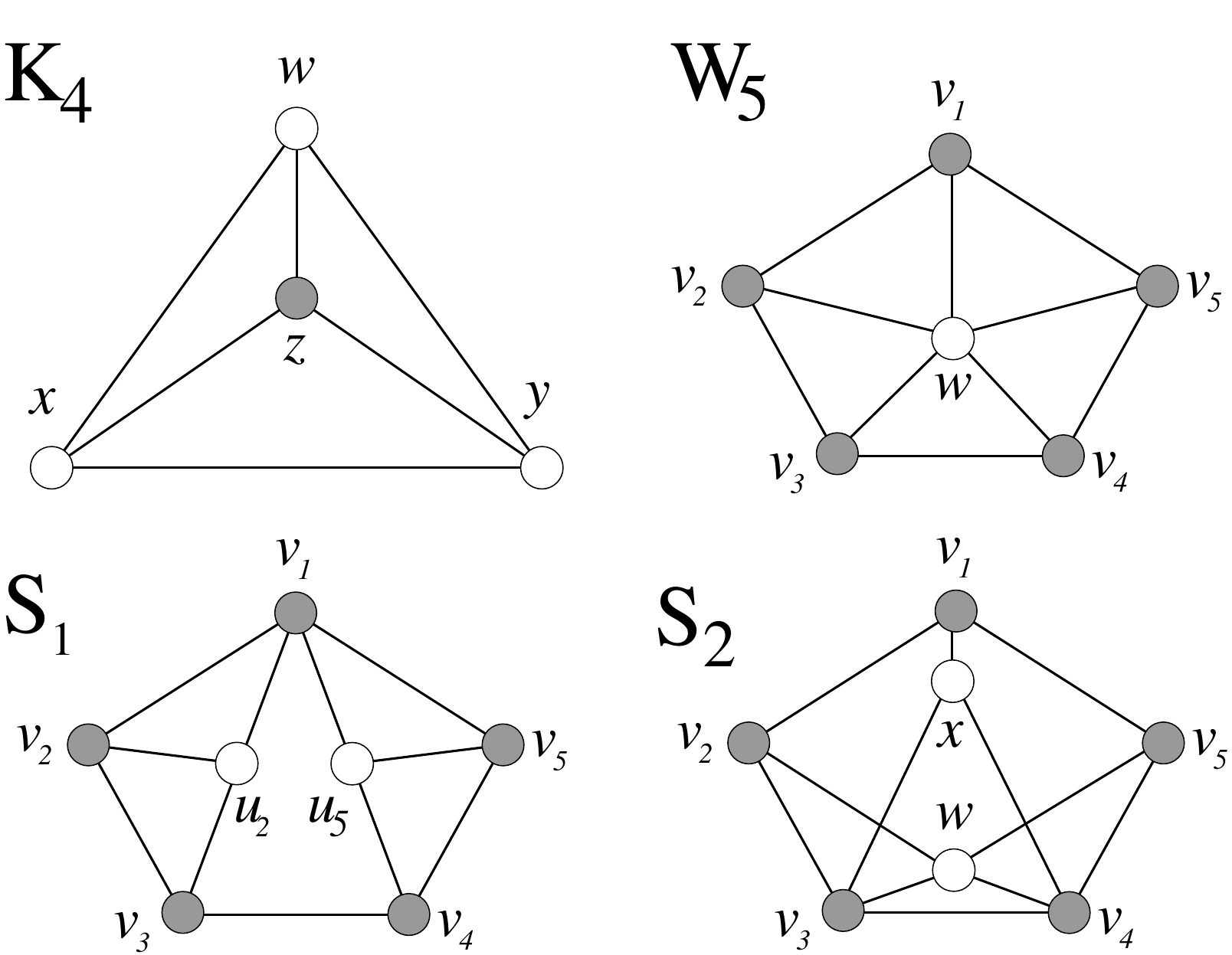}}
\end{minipage}
\hspace{0.5cm}
\begin{minipage}[b]{0.5\linewidth}
\centering \resizebox{!}{1.0in}{\includegraphics{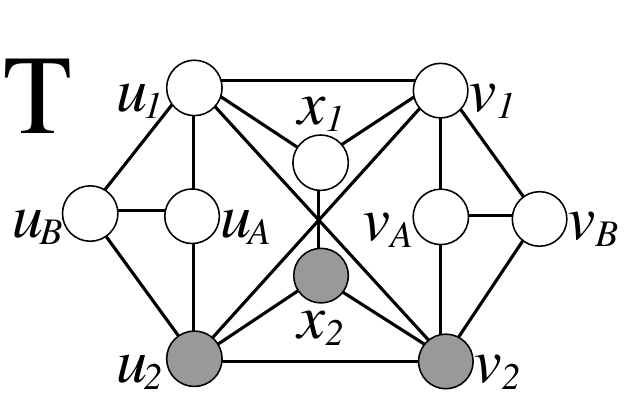}}
\centering \resizebox{!}{1.0in}{\includegraphics{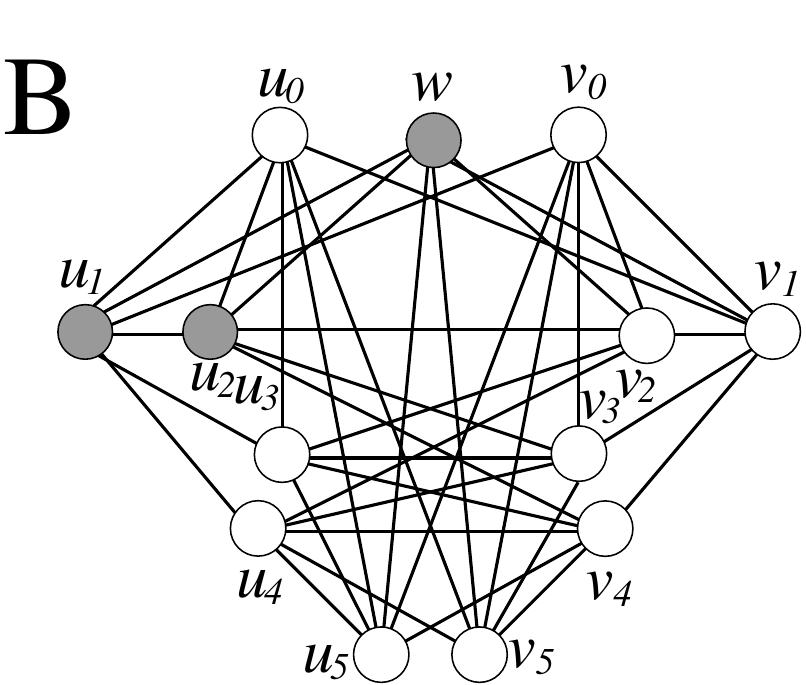}}
\end{minipage}

\caption{All 6 MN3P5s}\label{fig:6graphs}
\end{figure}

\begin{theorem}\label{thm:main}
A $P_5$-free graph is 3-colorable if and only if it does not
contain any of the six graphs in Fig.~\ref{fig:6graphs} as a
subgraph.
\end{theorem}
It is an easy matter to verify the graphs in
Fig.~\ref{fig:6graphs} are not 3-colorable, the rest of the
paper involves proving the other direction of the theorem. In the
last Section, we will discuss open problems arising from our work.

\section{Definition and Background}

Let $k$ and $t$ be positive integers. An MNkPt is a graph $G$ that
(i)  is not $k$-colorable and  is $P_t$-free and (ii) every proper
subgraph of $G$ is either $k$-colorable or has a $P_t$. We will be
interested specifically in the case where $k = 3$ and $t = 5$. We
will use the following notations. Let $G$ be a simple undirected
graph. A set $S$ of vertices of $G$ is {\it dominating} if every
vertex in $G-S$ has a neighbor in $S$. A $k$-clique is a clique on
$k$ vertices. $u \sim v$ will mean vertex $u$ is adjacent to
vertex $v$. $u \nsim v$ will mean vertex $u$ is not adjacent to
vertex $v$. For any vertex $v$, $N(v)$ denotes the set of vertices
that are adjacent to $v$. We write $G \cong H$ to mean $G$ is
isomorphic to $H$.  The {\it clique number} of $G$, denoted by
$\omega(G)$, is the number of vertices in a largest clique of $G$.
The {\it chromatic number } of $G$, denoted by $\chi(G)$, is the
smallest number of colors needed to color the vertices of $G$. A
{\it hole} is an induced cycle with at least four vertices, and it
is odd (or even) if it has odd (or even) length. An {\it
anti-hole} is the complement of a hole. A $k$-hole ($k$-anti-hole)
is a hole (anti-hole) on $k$ vertices. A graph $G$ is {\it
perfect} if each induced subgraph $H$ of $G$ has $\chi(G) =
\omega(G)$.
\begin{theorem}[The Strong Perfect Graph Theorem \cite{chu}]\label{thm:spgt} A graph is
perfect if and only if it does not contain an odd hole or odd
anti-hole as an induced subgraph.
\end{theorem}
Let  $\mathfrak{G} = \{K_4, W_5, S_1, S_2, T, B\}$ be the set of
graphs in Fig.~\ref{fig:6graphs}.  We will denote these graphs
in the following way.
\begin{itemize}
\item $P_5(v_1v_2v_3v_4v_5)$ means there is a $P_5$ being $v_1,
v_2, v_3, v_4$ and $v_5$.

\item $K_4(wxyz)$ means $\{w, x, y, z\}$ form a $K_4$.

\item $W_5(v_1v_2v_3v_4v_5, w)$ means $v_1, v_2, v_3, v_4, v_5$
and $w$ form a $W_5$ where $v_1v_2v_3v_4v_5$ form a 5-cycle and
$w$ is adjacent to every other vertex.

\item $S_1( v_1v_2v_3v_4v_5, u_2, u_5)$ means $v_1, v_2, v_3, v_4,
v_5, u_2, u_5$ form an $S_1$ where $v_1$ is the only degree 4
vertex and $N(v_1) = \{u_5, u_2, v_5, v_2\}$. Also $N(v_3) =
\{v_4, v_2, u_2\}$ and $N(v_4) = \{v_3, v_5, u_5\}$, and
$v_1v_2v_3v_4v_5$ form a 5-cycle.

\item $S_2(v_1v_2v_3v_4v_5, w, x)$ means $v_1, v_2, v_3, v_4, v_5,
w$ and $x$ form an $S_2$ where $N(w) = \{v_2, v_3, v_4, v_5\}$,
$N(x) = \{v_1, v_3, v_4\}$ and $v_1v_2v_3v_4v_5$ form a 5-cycle.

\item $T(u_1 u_A u_B u_2, v_1 v_A v_B v_2, x_1, x_2)$
 means a $T$ graph is present as shown previously.

\item $B(w, u_0u_1u_2u_3u_4u_5, v_0v_1v_2v_3v_4v_5)$   means a $B$ graph is
 present as shown previously.
\end{itemize}

We will rely on the following result.
\begin{theorem}[\cite{bacso}]\label{thm:bacso}
Every connected $P_5$-free graph has dominating clique or $P_3$.
\end{theorem}

The following lemma is folklore.

\begin{lemma}[The neighborhood lemma] \label{neighborhood}
Let $G$ be a minimally non $k$-colorable graph. If $u$ and $v$ are
two non-adjacent vertices in $G$, then $N(u) \nsubseteq N(v)$.

\end{lemma}

\begin{proof} Assume $N(u) \subseteq N(v)$.  Then the graph $G-v$
admits a $k$-coloring. By giving $u$ the color of $v$, we see that
$G$ is $k$-colorable, a contradiction.
\qed \end{proof}

The neighborhood lemma is used predominantly throughout this
paper. Writing $\mathbf{N(v, w) \rightarrow u}$ will denote the
fact that $N(v) \nsubseteq N(w)$ by the neighborhood lemma so
there exists a vertex $u$ where $u \sim v$, but $u \nsim w$.

The following fact is well-known and easy to establish.

\begin{fact}\label{degree 3 theorem}

In a minimally non $k$-colorable graph every vertex has degree at
least $k$. $\Box$

\end{fact}

\section{Intermediate Results}
In this section, we establish a number of intermediate results
needed for proving the main theorem.

\begin{lemma}\label{cycle degree 4 lemma}

Let $G$ be an MN3P5 graph with a 5-hole $C = \{v_1, v_2, v_3, v_4,
v_5\}$ and a vertex $w$ adjacent to at least 4 vertices of $C$.
Then $G \in \mathfrak{G}$.

\end{lemma}

\begin{proof} If $w$ is adjacent to all five vertices of $C$, then $G$ clearly is isomorphic to
$W_5$. Now, assume $N(w) \cap \{v_1, v_2, v_3, v_4, v_5\} = \{v_2,
v_3, v_4, v_5\}$.

We have $\mathbf{N(v_1, w)\rightarrow x}$.

Assume for the moment that $x \nsim v_3, v_4$. We have
\newline
\hspace*{3em} $x
\sim v_5$, otherwise,  we have $P_5(xv_1v_5v_4v_3)$. \\
\hspace*{3em} $x \sim v_2$, otherwise, we have
$P_5(xv_1v_2v_3v_4)$.

But then $G$ contains $S_1(v_1v_2v_3v_4v_5, x, w)$. This means $x
\sim v_3$ or $x \sim v_4$. By symmetry, we may assume $x \sim
v_3$. We have $x \sim v_2$ or $x \sim v_4$, otherwise, $G$
contains $P_5(xv_1v_2wv_4)$. If $x \sim v_2$  then $G$ properly
contains $S_1(v_1v_2v_3v_4v_5, x, w)$, a contradiction. This means
$x \sim v_4$; so $G$ contains $S_2(v_1v_2v_3v_4v_5, w, x)$ and $G
\cong S_2$. \qed
\end{proof}
\begin{theorem}\label{thm:5-hole}

Every MN3P5 graph different from $K_4$ contains a 5-hole.

\end{theorem}

\begin{proof}
Let $G$ be an MN3P5 graph different from a $K_4$. We have
$\omega(G) \leq 3$ and $\chi(G) \geq 4$. Thus, $G$ is not perfect.
By Theorem \ref{thm:spgt}, $G$ contains an odd hole or an odd
anti-hole $H$. $H$ cannot be a hole of size 7 or greater because
$G$ is $P_5$-free. We may assume $H$ is an anti-hole of length at
least seven, for otherwise we are done (observe that the hole on
five vertices is self-complementary). Let $v_1$, $v_2$, $v_3$,
$v_4$, $v_5$, $v_6$, $v_7$ be the cyclic order of the hole in the
complement of $G$. Then $G$ properly contains $S_1(v_4 v_6 v_3 v_5
v_2, v_1, v_7)$, a contradiction.
\qed \end{proof}

\begin{lemma}\label{lem:k3}

Let $G$ be an MN3P5 graph that has a dominating clique $\{a, b,
c\}$. Also assume that there is a vertex $v \notin \{a, b, c\}$
adjacent to two vertices from $\{a, b, c\}$. Then $G \in
\mathfrak{G}$.

\end{lemma}
\begin{proof} The proof is by contradiction. Suppose that $G \notin
\mathfrak{G}$. We may assume $v$ is adjacent to $b$ and $c$. We
have $v \nsim a$, otherwise, $G$ contains $K_4(abcv)$. Through
repeated applications of the Neighborhood Lemma, we will
eventually add nine vertices to $G$ to arrive at a contradiction.
In the end, we will obtain the graph B (see Fig.~\ref{fig:Bgraph} for the order in which vertices are added). Each
time we add a vertex we will consider its adjacency to the other
vertices of the graph. In every case, the adjacency can be
completely determined at each step.
\begin{figure}[ht]

\centering \resizebox{2in}{!}{\includegraphics{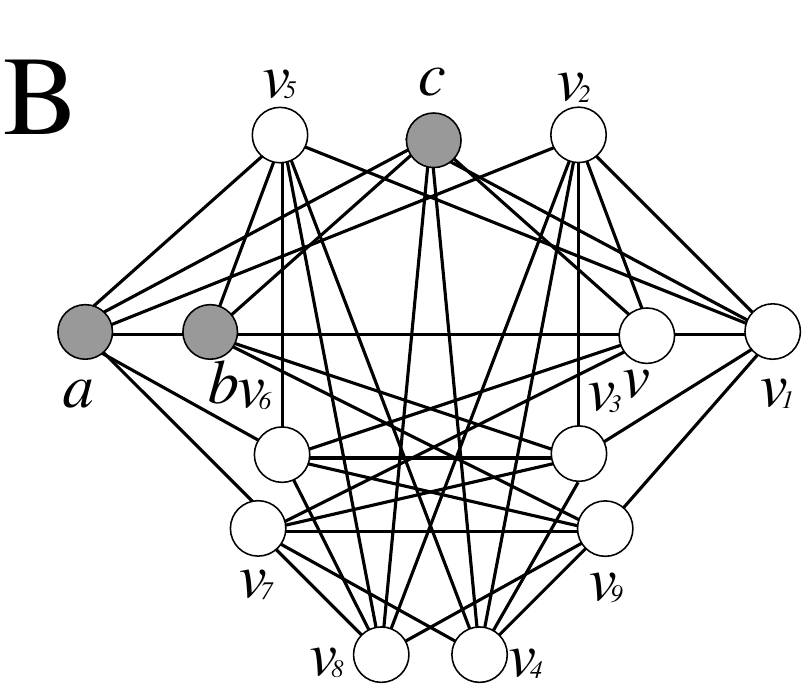}}

\caption{The graph B obtained in the proof of
Lemma~\ref{lem:k3}}\label{fig:Bgraph}
\end{figure}

\noindent $\mathbf{N(v, a) \rightarrow v_1}$.
\begin{myparagraph}

\item $v_1 \sim c$: since $\{a,b,c\}$ is dominating, $v_1$ is
adjacent to either $b$ or $c$. Without loss of generality, assume
$v_1 \sim c$.

\item $v_1 \nsim b$: otherwise, $G$ contains $K_4(bcvv_1)$.

\end{myparagraph}
$\mathbf{N(v_1, b) \rightarrow v_2}$.

\begin{myparagraph}

\item $v_2 \sim a$: assume $v_2 \nsim a$. We have $v_2 \sim v$,
otherwise, $G$ contains $P_5(v_2 v_1 v b a)$. Also, $v_2 \sim c$
since  $\{a, b, c\}$ is a dominating set. But then, $G$ contains
$K_4(v_1v_2vc)$.

\item $v_2 \nsim c$: otherwise, $G$ contains $W_5(abv v_1 v_2,
c)$.

\item $v_2 \sim v$: otherwise, $c$ has four neighbors in the
5-hole $v_2 a b v v_1$  contradicting Lemma \ref{cycle degree 4
lemma}.

\end{myparagraph}
$\mathbf{N(v_2, c)\rightarrow v_3}$.

\begin{myparagraph}

\item $v_3 \sim b$: assume $v_3 \nsim b$. We have $v_3 \sim a$
since $\{a, b, c\}$ is a dominating set. We have $v_3 \nsim v_1$,
otherwise, $G$ contains $S_1(v b a v_3 v_2, c, v_1)$.  But then
$G$ contains $P_5(v_3v_2v_1cb)$.

\item $v_3 \nsim v$; otherwise, $G$ contains $W_5(bcv_1v_2v_3,
v)$.

\item $v_3 \sim v_1$: otherwise, $v$ has four neighbors in the
5-hole $v_3bcv_1v_2$  contradicting Lemma \ref{cycle degree 4
lemma}.

\item $v_3 \nsim a$: otherwise $G$ contains $S_1(v_3 a c v v_1, b,
v_2)$.

\end{myparagraph}
$\mathbf{N(v_3, v)\rightarrow v_4}$.

\begin{myparagraph}

\item $v_4 \sim c$: assume $v_4 \nsim c$. Then we have $v_4 \sim
v_2$, for otherwise $G$ contains $P_5(v_4v_3v_2vc)$; $v_4 \nsim
v_1$, for otherwise $G$ contains $K_4(v_1v_2v_3v_4)$;  $v_4 \nsim
b$, for otherwise $G$ contains $S_1(v_1 v_2 v_4 b c, v_3, v)$;
$v_4 \sim a$ because $\{a,b,c\}$ is dominating. But then $G$
contains $P_5(v_4abvv_1)$.

\item $v_4 \nsim v_1$: for otherwise $G$ contains
$W_5(v_4v_3v_2vc, v_1)$.

\item $v_4 \nsim b$: for otherwise, $G$ contains $S_1(v v_1 v_3
v_4 b, v_2,c)$.

\item $v_4 \nsim a$: for otherwise, $G$ contains $P_5(v_4abvv_1)$.

\item $v_4 \sim v_2$: for otherwise the vertex $v_1$ has exactly
four neighbors in the 5-hole $v_4v_3v_2vc$ contradicting Lemma
\ref{cycle degree 4 lemma}.

\end{myparagraph}
$\mathbf{N(a, v)\rightarrow v_5}$.

\begin{myparagraph}

\item $v_5 \nsim v_3$: Assume $v_5 \sim v_3$. Then we have $v_5
\sim v_1$, for otherwise $G$ contains $P_5(av_5v_3v_1v)$; $v_5
\nsim v_2$, for otherwise $G$ contains $K_4(v_1v_2v_3v_5)$; $v_5
\sim c$, for otherwise $G$ contains $P_5(v_5 v_3 v_2 v c)$. But
now $G$ contains $W_5(v_5 c v v_2 v_3, v_1)$.

\item $v_5 \sim b$: assume $v_5 \nsim b$. Then we have $v_5 \sim
v_1$, for otherwise $G$ contains $P_5(v_5abvv_1)$. But then $c$
has four neighbors in the 5-hole $v_5abvv_1$ contradicting Lemma
\ref{cycle degree 4 lemma}.

\item $v_5 \nsim c$: for otherwise $G$ contains $K_4(abcv_5)$.

\item $v_5 \sim v_1$: for otherwise $G$ contains
$P_5(v_5acv_1v_3)$.

\item $v_5 \sim v_4$: for otherwise $G$ contains
$P_5(v_3v_4cav_5)$.

\item $v_5 \nsim v_2$: for otherwise $G$ contains $S_1(c v v_2 v_5
a, v_1, b)$.

\end{myparagraph}
$\mathbf{N(v_5, c)\rightarrow v_6}$.

\begin{myparagraph}

\item $v_6 \sim v$: assume $v_6 \nsim v$. We have $v_6 \sim a$,
for otherwise $G$ contains $P_5(v_6v_5acv)$; $v_6 \nsim b$, for
otherwise $G$ contains $K_4(abv_5v_6)$;  $v_6 \sim v_1$, for
otherwise, $G$ contains $P_5(v_6abvv_1)$. But $c$ has four
neighbors in the 5-hole $v_6abvv_1$ contradicting Lemma \ref{cycle
degree 4 lemma}.

\item $v_6 \nsim b$: for otherwise $G$ contains $W_5(v_5v_6vca,
b)$.

\item $v_6 \nsim v_2$: for otherwise $G$ contains
$P_5(v_2v_6v_5bc)$.

\item $v_6 \sim v_3$: for otherwise $G$ contains
$P_5(v_5v_6vv_2v_3)$

\item $v_6 \sim a$: for otherwise $G$ contains
$P_5(v_3v_6v_5ac)$.

\item $v_6 \nsim v_1$: for otherwise $G$ contains  $S_1(v_6 a b c
v, v_5, v_1)$.

\item $v_6 \nsim v_4$: for otherwise $G$ contains  $T(v_6av_5b,
v_3v_2v_1v, v_4, c)$.

\end{myparagraph}
$\mathbf{N(v_4, v_1)\rightarrow v_7}$.

\begin{myparagraph}
\item $v_7 \sim v$: assume $v_7 \nsim v$. Then we have $v_7 \sim
v_3$, for otherwise $G$ contains $P_5(v_7v_4v_3v_1v)$; $v_7 \nsim
v_2$, for otherwise $G$ contains  $K_4(v_2v_3v_4v_7)$; $v_7 \sim
c$, for otherwise $G$ contains  $P_5(v_7v_3v_2vc)$. Now,  $G$
contains $S_1(v_2 v c v_7 v_3, v_1, v_4)$.

\item $v_7 \nsim v_2$: for otherwise $G$ contains
$W_5(vv_1v_3v_4v_7, v_2)$.

\item $v_7 \nsim v_6$: for otherwise $G$ contains
$P_5(v_6v_7v_4v_2v_1)$.

\item $v_7 \sim a$: for otherwise $G$ contains $P_5(v_4v_7vv_6a)$.

\item $v_7 \sim v_3$: for otherwise $G$ contains
$P_5(av_7vv_1v_3)$.

\item $v_7 \nsim c$: for otherwise $G$ contains $S_1(v_3 v_4 c v
v_1, v_7, v_2 )$.

\item $v_7 \nsim b$: for otherwise $G$ contains
$P_5(v_7bcv_1v_2)$.

\item $v_7 \nsim v_5$: for otherwise $G$ contains $T(av_7v_5v_4,
cvv_1v_2, b, v_3)$.

\end{myparagraph}
$\mathbf{N(v_6, b)\rightarrow v_8}$.

\begin{myparagraph}

\item $v_8 \sim c$: assume $v_8 \nsim c$. Then we have $v_8 \sim
a$ because $\{a, b, c\}$ is a dominating set; $v_8 \sim v_5$, for
otherwise $G$ contains $P_5(v_8v_6v_5bc)$. But now, $G$ contains
$K_4(av_5v_6v_8)$.

\item $v_8 \nsim a$: for otherwise $G$ contains $W_5(v_8 v_6 v_5
bc, a)$.

\item $v_8 \nsim v_1$: for otherwise $G$ contains
$P_5(bav_6v_8v_1)$.

\item $v_8 \sim v_2$: for otherwise $G$ contains
$P_5(v_2v_1cv_8v_6)$.

\item $v_8 \sim v_5$: for otherwise $G$ contains
$P_5(v_8v_2v_1v_5b)$.

\item $v_8 \nsim v_4$: for otherwise $G$ contains
$P_5(v_4v_8v_6ab)$.

\item $v_8 \nsim v$: for otherwise $G$ contains  $S_1(b c v_8 v_6
a, v, v_5)$.

\item $v_8 \nsim v_3$: for otherwise $G$ contains  $T(v_6av_5b,
v_3v_2v_1v, v_8, c)$.

\item $v_8 \sim v_7$: for otherwise $G$ contains
$P_5(v_8v_5bv_3v_7)$.

\end{myparagraph}
$\mathbf{N(v_8, a)\rightarrow v_9}$.

\begin{myparagraph}

\item $v_9 \sim b$:  assume $v_9 \nsim b$. We have $v_9 \sim v_2$,
for  otherwise $G$ contains $P_5(v_9v_8v_2ab)$; $v_9 \sim v_6$,
for  otherwise $G$ contains  $P_5(v_9v_8v_6ab)$. This means $G$
contains $T(v_6v_8v_9v_2, abcv, v_5, v_1)$.

\item $v_9 \sim v_1$: assume $v_9 \nsim v_1$. We have $v_9 \sim
v_2$, for  otherwise $G$ contains  $P_5(v_9bav_2v_1)$. This means
$G$ contains $T(v_2vv_1c, v_8v_6v_5a, v_9, b)$.

\item $v_9 \sim v_6$: for  otherwise $G$ contains
$P_5(v_1v_9bav_6)$.

\item $v_9 \sim v_7$: for  otherwise $G$ contains
$P_5(v_1v_9bav_7)$.

\item $v_9 \sim v_4$: assume $v_9 \nsim v_4$. Then we have $v_9
\sim v_2$, for  otherwise $G$ contains  $P_5(v_9bav_2v_4)$. This
means $G$ contains $T(v_6av_5b, v_9v_2v_1v, v_8, c)$.

\end{myparagraph}

But this means $G$ contains $B(c, v_5 a b v_6 v_7 v_8, v_2 v_1 v
v_3 v_9 v_4)$, a contradiction.
\qed \end{proof}
\begin{lemma}\label{lemma D4}

Let $G$ be an MN3P5 with a dominating clique $\{a, b, c\}$. Let
$A$ = $N(a) - \{b, c\}$, $B = N(b) - \{a, c\}$ and $C = N(c) -
\{a, b\}$. Suppose $A$, $B$ and $C$ are pairwise disjoint. Then $G
\in \mathfrak{G}$.

\end{lemma}

\begin{proof} Some observations are necessary for this proof.
\begin{observation}\label{obs:component}
Let $X$ and $Y$ be two distinct elements of $\{A,B,C\}$. Let $X'$
be a component in $X$ with at least two vertices, and $y$ be a
vertex in $Y$. Then either $y$ is adjacent to all vertices of $X'$
or to no vertex of $X'$.
\end{observation}
\begin{proof} Suppose the Observation is false. Then there are
adjacent vertices $v_1, v_2 \in X$ such that $y$ is adjacent to
exactly one of $v_1, v_2$. Without loss of generality, we may
assume $X=A$ and $Y=B$. Now, $\{c, b, y, v_2, v_1\}$ induces a
$P_5$, a contradiction.
\qed \end{proof}
\begin{observation}\label{obs:edge}
Every component in $A$, $B$ or $C$ is a single edge or one vertex.
\end{observation}
\begin{proof}
Assume that one of $A$, $B$ or $C$ contains a vertex of degree 2.
Without loss of generality, assume there is such a vertex $a_0 \in
A$ that is adjacent to two other distinct vertices $a_1$ and $a_x$
in $A$. Now we have $a_1 \nsim a_x$, for otherwise $G$ contains
$K_4(a_1a_xa_0a)$. The Neighborhood Lemma implies $\mathbf{N(a_1,
a_x)\rightarrow a_2}$ and $\mathbf{N(a_x, a_1)\rightarrow a_y}$.
Observation \ref{obs:component} implies $a_2, a_y \in A$. We have
$a_y \nsim a_0$, for otherwise $G$ contains $K_4(aa_0a_xa_y)$;
$a_2 \nsim a_0$,  for otherwise $G$ contains $K_4(aa_0a_1a_2)$;
$a_y \sim a_2$,  for otherwise $G$ contains
$P_5(a_ya_xa_0a_1a_2)$. Then $G$ contains $W_5(a_ya_xa_0a_1a_2,
a)$, a contradiction.
\qed \end{proof}

We continue the proof of the Lemma. Assume $G \notin
\mathfrak{G}$. Consider the case that two of $A$, $B$ or $C$
contain an edge. Without loss of generality, assume $A$ contains
an edge $a_1 a_2$ and $B$ contains an edge $b_1 b_2$. If a vertex
in $\{b_1, b_2\}$ is adjacent to a vertex in $\{a_1, a_2\}$ then
by Observation \ref{obs:component}, $G$ contains
$K_4(a_1a_2b_1b_2)$, a contradiction. Suppose some vertex $c_0 \in
C$ is adjacent to a vertex in $\{a_1, a_2, b_1, b_2\}$. We may
assume $c_0 \sim a_1$. By Observation \ref{obs:component}, we have
$c_0 \sim a_2$. If $c_0 \nsim b_i$ ($i=1,2$) then $G$ contains
$P_5 (b_i b c c_0 a_1)$. So, $c_0$ is adjacent to all vertices of
$\{a_1, a_2, b_1, b_2\}$. But now, $G$ contains $S_1 (c_0 a_1 a b
b_1, a_2, b_2)$. So, no vertex in $C$ is adjacent to a vertex in
$\{a_1, a_2, b_1, b_2\}$. By Fact \ref{degree 3 theorem} and
Observation \ref{obs:edge}, there exists a vertex $a_3 \in A$ with
$b_1, b_2 \sim a_3$ and a vertex $b_3 \in B$ with $a_1, a_2 \sim
b_3$. Also by Fact \ref{degree 3 theorem}, $C$ contains a vertex
$c_0$. We have $a_3 \sim c_0$, for otherwise $G$ contains
$P_5(a_3b_1bcc_0)$; $b_3 \sim c_0$, for otherwise $G$ contains
$P_5(b_3a_1acc_0)$; $a_3 \sim b_3$, for otherwise $G$ contains
$P_5(b_1 a_3 c_0 b_3 a_1)$. But now $G$ contains $T(aa_1a_2b_3,
bb_1b_2a_3, c, c_0)$ which is a contradiction. So, at most one of
$A,B,C$ contains an edge.

If all of $A,B,C$ is a stable set, then $G$ is obviously
3-colorable. We may assume $B,C$ are stable sets, and $A$ contains
an edge. Now there must be one vertex $b_0 \in B$ with $N(b_0)$
contains two adjacent vertices in $A$. Otherwise, $G$ admits a
3-coloring $f$ as follows. The vertices of $C$ are colored with
color 3. Now, for each edge in $A$, its endpoints are arbitrarily
colored with colors 1, 2. The remaining vertices of $A$ are
colored with color 1. The vertices of $B$ are colored with color 2
(no vertex of $B$ is adjacent to an endpoint of a edge of $A$ by
Observation \ref{obs:component}), and let $ f(a) = 3, f(b) = 1,
f(c) = 2$. Thus, $f$ is a 3-coloring which is a contradiction.
Therefore, there is a vertex $b_1 \in B$ adjacent to both
endpoints in some edge $a_{b1} a_{b2}$ in $A$. By a similar
argument, there is a vertex $c_1 \in C$ adjacent to both endpoints
in some edge $a_{c1} a_{c2}$.

Suppose that $a_{b1} a_{b2} $ and $a_{c1} a_{c2}$ are the same
edge. For simplicity, write $a_1 a_2 = a_{b1} a_{b2}= a_{c1}
a_{c2}$. We have $b_1 \nsim c_1$, for otherwise $G$ contains
$K_4(a_1a_2b_1c_1)$.

\begin{myparagraph}

\item $\mathbf{N(b_1, a)\rightarrow c_2}$. We have $c_2 \in C$ by
the fact that $B$ is an independent set.

\item $\mathbf{N(c_1, a)\rightarrow b_2}$. We have $b_2 \in B$ by
the fact that $C$ is an independent set.

\item $b_2, c_2 \nsim a_1, a_2$. Otherwise, suppose $b_2 \sim
a_1$. Then by Observation~\ref{obs:component}, we have $b_2 \sim
a_2$ so $G$ contains $K_4(a_1a_2b_2c_1)$.

\item $b_2 \sim c_2$. Otherwise, $G$ contains $P_5(c_1 b_2 b b_1
c_2)$.


\end{myparagraph}

Now, $G$ contains $P_5(b_2 c_2 c a a_1)$. Thus, $a_{b1} a_{b2} $
and $a_{c1} a_{c2}$ are distinct edges. We have $b_1 \nsim a_{c1},
a_{c2}$ and $c_1 \nsim a_{b1}, a_{b2}$, for otherwise we are done
by the previous case. We have $b_1 \sim c_1$, for otherwise $G$
contains $P_5(b_1 a_{b1} a a_{c1} c_1)$. But now $G$ contains
$S_1(ab_1a_{b1} b_1 c_1 a_{c1}, a_{b2}, a_{c2})$, a contradiction.
\qed \end{proof}

\begin{lemma}\label{lemma D5}

Let $G$ be an MN3P5 with a dominating clique $\{a, b, c\}$. Then
$G \in \mathfrak{G}$.

\end{lemma}

\begin{proof}
If there is a vertex other than $a$, $b$ and $c$ adjacent to at
least two of $a, b$ or $c$ then by Lemma \ref{lem:k3}, $G \in
\mathfrak{G}$. Otherwise, the conclusion follows from Lemma
\ref{lemma D4}.
\qed \end{proof}
\begin{lemma}\label{lemma D6}

Let $G$ be an MN3P5 with a dominating clique $\{a, b\}$ of size 2.
Then $G \in \mathfrak{G}$.

\end{lemma}

\begin{proof}
Assume $G \notin \mathfrak{G}$. We may assume $G$ contains no
dominating 3-clique, for otherwise we are done by Lemma \ref{lemma
D5}. It follows  that no vertex $v$ is adjacent to both $a,b$.

By Theorem \ref{thm:5-hole}, there is 5-hole $C = v_1 v_2 v_3 v_4
v_5$ in $G$ because $G \neq K_4$. Clearly $C$ cannot contain both
$a$ and $b$.  WLOG, assume that $|N(a) \cap C| \geq |N(b) \cap
C|$. If $b \notin C$ then since $\{a,b\}$ is a dominating clique
of $G$ we have $|N(a) \cap C| \geq 3$. If $b \in C$, then $a$ must
be adjacent to the 2 vertices in $C$ not adjacent to $b$. Thus,
since $a \sim b$ we also have $|N(a) \cap C| \geq 3$. The case
when $|N(a) \cap C| \geq 4$ is handled by Lemma \ref{cycle degree
4 lemma}, so WLOG we may assume either $N(a) \cap C =\{v_1, v_2,
v_3\}$ or   $N(a) \cap C =\{v_1, v_3, v_4\}$.


Suppose $N(a) \cap C =\{v_1, v_2, v_3\}$. Since $\{a,b\}$ is a
dominating clique, we have $b \not\in C$ and $b \sim v_4, v_5$.
Since no vertex is adjacent to both $a$ and $b$, $G$ contains $P_5
(b v_5 v_1 v_2 v_3)$, a contradiction. Now, we may assume  $N(a)
\cap C =\{v_1, v_3, v_4\}$. There exists a vertex $x$ with $x
\nsim a, v_3, v_4$, for otherwise $\{a, v_3, v_4\}$ is dominating
3-clique. If $x \sim v_5$, then $x \sim v_2$, for otherwise $G$
contains $P_5 (x v_5 v_4 v_3 v_2)$; but now $G$ contains $P_5 (v_2
x v_5 v_4 a)$. Thus, we have $x \nsim v_5$ and by symmetry $x
\nsim v_2$. Since $\{a,b\}$ is a dominating clique, we have $x
\sim b$, and $b \sim v_2, v_5$. Recall that no vertex is adjacent
to both $a,b$. Now, $G$ contains $P_5 (x b v_5 v_4 v_3) which is a contradiction.$
\qed \end{proof}

\begin{theorem}\label{dominating clique theorem}

If $G$ is an MN3P5 with a dominating clique then $G \in
\mathfrak{G}$.

\end{theorem}

\begin{proof}
If $G$ has a dominating clique of size one or two, then it has a
dominating clique of size 2 since $G$ contains no isolated
vertices. By Lemma \ref{lemma D6}, $G \in \mathfrak{G}$. If $G$
has a dominating clique of size 3, then Lemma \ref{lemma D5}
implies $G \in \mathfrak{G}$. If $G$ has a dominating clique of
size 4 or more, then $G$ contains a $K_4$ so $G = K_4 \in
\mathfrak{G}$ by minimality.
\qed \end{proof}

\begin{lemma}\label{dominating C5 theorem}

Let $G$ be an MN3P5 with a dominating 5-hole. Then $G$ has a
dominating $K_3$ or  $G \in \mathfrak{G}$.

\end{lemma}
\begin{proof}
Let $C= v_1 v_2 v_3 v_4 v_5$ be an induced 5-hole of $G$. Assume
$G$ does not have a dominating clique. Let $X_i$ be the set of
vertices adjacent to $v_{i-1}$ and $v_{i+1}$ and not adjacent to
$v_{i+2}$ and $v_{i+3}$ with the subscript taken modulo $5$, for
$i=1,2,3,4,5$. We now prove every vertex of $G$ belongs to exactly
one $X_i$.

Consider a vertex $w \not\in C$. By Lemma \ref{cycle degree 4
lemma}, we have $1 \leq |N(w) \cap C| \leq 3$. If $w$ has one
neighbor in $C$, then $G$ obviously contains a $P_5$. Suppose $w$
has two neighbors $a,b$ in $C$. If $a \sim b$, then $G$ obviously
contains a $P_5$. Otherwise, $a$ and $b$ have distance two on $C$
and so $w$ belongs to some $X_i$. We may now assume $w$ has three
neighbors on $C$. If these three neighbors are consecutive on $C$,
then $w$ belongs to some $X_i$. Now, we may assume $w \sim v_1,
v_3, v_4$. There is a vertex $x$ with $x \nsim w, v_4, v_3$, for
otherwise $\{w, v_4, v_3\}$ is a dominating clique. Vertex $x$
must have a neighbor in $\{v_1, v_2, v_5\}$ because $C$ is a
dominating set. If $x \sim v_5$, then $x \sim v_2$, for otherwise
$G$ contains $P_5 (x v_5 v_4 v_3 v_2)$; but now $G$ contains $P_5
(v_2 x v_5 v_4 w)$. Thus, we have $x \nsim v_5$ and by symmetry $x
\nsim v_2$. Now, we have $x \sim v_1$, and $G$ contains $P_5 (x
v_1 v_5 v_4 v_3)$. Thus, $X_1, X_2, X_3, X_4, X_5$ is a partition
of $V(G)$.

If there are nonadjacent vertices $x_1, x_2$ with $x_1 \in X_1,
x_2 \in X_2$, then $G$ contains $P_5 (x_1 v_5 v_4 v_3 x_2)$. Thus,
there are all possible edges between $X_i$ and $X_{i+1}$ for all
$i$. If every $X_i$ is a stable set, then $G$ is obviously
3-colorable, a contradiction. So we may assume WLOG $X_5$ contains
an edge $ab$. Then $X_1$ is a stable set, for otherwise $G$
contains a $K_4$ with one edge in $X_1$ and one edge in $X_5$.
Similarly, $X_4$ is a stable set. If $X_2$ contains an edge $cd$,
then $G$ contains $S_1 (v_1 c v_3 v_4 a, d, b)$. If $X_3$ contains
an edge $fg$, then $G$ contains $S_1 (v_4 f v_2 v_1 a, g,b)$.
Thus, $X_i$ is a stable set for $i=1,2,3,4$. Consider the subgraph
$H$ of $G$ induced by $X_5$. If $H$ contains an odd cycle $D$,
then $D \cup \{v_1\}$ is a $K_4$ or $W_5$, or $D$ contains a
$P_5$. Thus $H$ is bipartite. By coloring $X_5$ with colors 2,3,
$X_1 \cup X_4$ with color 1, $X_2$ with color 2, $X_3$ with color
3, we see that $G$ is 3-colorable, a contradiction.
\qed \end{proof}
\section{Proof of Theorem~\ref{thm:main}}
We can now prove the main theorem.

It is a routine matter to verify  the ``only if'' part. We only
need prove the ``if'' part. Suppose $G$ does not contain any of
the graphs in Fig.~\ref{fig:6graphs} but is not 3-colorable.
Then $G$ contains an induced subgraph that is minimally not
3-colorable. It follows that we may assume $G$ is a connected
MN3P5 graphs. By Theorem~\ref{thm:bacso}, $G$ contains  a
dominating clique or $P_3$. If $G$ contains a dominating clique,
then we are done by Theorem~\ref{dominating clique theorem}. So,
we may assume $G$ contains no dominating clique and thus contains
a dominating $P_3$ with vertices  $v_1, v_2, v_3$ and edges $v_1
v_2, v_2 v_3$. There is a vertex $v_4$ with $v_4 \sim v_3$ and
$v_4 \nsim v_1, v_2$ since $v_1 v_2$ is not a dominating edge.
Similarly, there is a vertex $v_5$ with $v_5 \sim v_1$ and $v_5
\nsim v_2, v_3$. We have $v_5 \sim v_4$, for otherwise $G$
contains a $P_5$. Thus, $v_1 v_2 v_3 v_4 v_5$ is a dominating
5-hole of $G$, and we are done by Lemma~\ref{dominating C5
theorem}. \hfill $\Box$
\section{Conclusion and Open Problems}\label{sec:last}
In this paper, we provide a certifying algorithm for the problem
of 3-coloring a $P_5$-graph by showing there are exactly six
finite minimally non-3-colorable graphs. Previously known
algorithms (\cite{Hoa2,mel,woe}) provide a yes-certificate by
constructing a 3-coloring if one exists. Our algorithm provides a
no-certificate by finding one of the six graphs of
Fig.~\ref{fig:6graphs}. Since these graphs are finite, our
algorithm runs in polynomial time. We do not know if there is a
fast algorithm running in, say, $O(n^4)$ to test if a graph
contains one of the six graphs of Fig.~\ref{fig:6graphs} as a
subgraph. We leave this as an open problem.

In \cite{Hoa,Hoa2}, it is shown for every fixed $k$, determining
if a $P_5$-free graph is $k$-colorable is polynomial-time
solvable. It is tempting to speculate that these two algorithms
work because for every fixed $k$, there is a function $f(k)$ such
that every minimally non-$k$-colorable $P_5$-free graph has at
most $f(k)$ vertices. The result of this paper can be viewed as a
first step in this direction.



\begin{thebibliography}{99}




\bibitem{bacso}
G. Bacs\'{o} and Z. Tuza, Dominating cliques in $P_5$-free graphs,
Period. Math. Hungar. Vol. 21 No. 4 (1990) 303-308.



\bibitem{cop}
D. Coppersmith and S. Winograd, Matrix multiplication via
arithmetic progressions. Journal of Symbolic Computation, Vol. 9
No. 3(1990) 251-280.

\bibitem{chu}
M. Chudnovsky, N. Robertson, P. Seymour, R. Thomas. The strong
perfect graph theorem. Annals of Mathematics 164 (1) (2006):
51–-229.















\bibitem{Hoa}
C. T. Ho\`ang, M. Kamin\'ski, V. Lozin, J. Sawada, X. Shu,
Deciding k-colorability of P5-free graphs in polynomial time, to
appear in Algorithmica.

\bibitem{Hoa2}
C. T. Ho\`ang, M. Kamin\'ski, V. Lozin, J. Sawada, X. Shu, A Note
on k-Colorability of P5-Free Graphs, Lecture Notes In Computer
Science; Vol. 5162 (2008) 387 - 394.





\bibitem{Kor90}
D.V.~Korobitsyn, On the complexity of determining the domination
number in monogenic classes of graphs, Diskret. Mat. 2, N 3
(1990), 90-96 in Russian, translation in  Discrete Mathematics and
Applications, 2 (1992), no. 2, 191-199.

\bibitem{kral}
D. Kral, J. Kratochvil, Z. Tuza and G. J. Woeginger, Complexity of
coloring graphs without forbidden induced subgraphs, in: WG 2001,
LNCS 2204, (2001) 254-262.

\bibitem{KraMcc}
D. Kratsch, R. M. McConnell, K. Mehlhorn, and J. P. Spinrad.
Certifying algorithms for recognizing interval graphs and
permutation graphs. SIAM J. Comput., 36(2):326–-353, 2006.



\bibitem{LRS06} V. Bang Le, B. Randerath, I. Schiermeyer,
On the complexity of 4-coloring graphs without long induced
paths, Theoretical Computer Science 389 (2007) 330–-335.


\bibitem{mel}
S. Mellin, Polynomielle F\"arbungsalgorithmen f\"ur $P_k$-freie
Graphen, Diplomarbeit am Institut f\"ur Informatik, Universit\"at
zu K\"oln, (2002).

\bibitem{RS04}  B. Randerath, I. Schiermeyer, Vertex coloring and forbidden subgraphs -- a
survey, Graphs and Combinatorics, 20(1) (2004) 1-40.

\bibitem{RS04A} B. Randerath, I. Schiermeyer, $3$-colorability
$\in \mathcal{P}$ for $P_6$-free graphs, Discrete Applied
Mathematics 136 (2004) 299-313.



\bibitem{woe} G.~J.~Woeginger, J. Sgall, The complexity of coloring graphs without long induced paths, Acta Cybernetica 15(1),
(2001) 107-117.

\end{thebibliography}
\end{document}